# Zitterbewegung and the Electron

Arend Niehaus*

## Abstract

Starting from a statistical model of the electron, which explains spin and spin measurements in terms of a probability density distribution resulting from a rapidly changing angular momentum during an extended Zitterbewegung (EZBW), a "light-like" model of the electron and other spin-1/2 particles is formulated. This model describes individual particles in terms of paths of a moving quantum. It is shown that this description allows one to reproduce observable properties as path-averages over a period of the fast (EZBW) in elementary calculations. The general topology of the paths may be described as a helical path, with a helix axis forming a circle around a fixed point in space. The radius of the helix and of the circle are equal and given by half the reduced Compton wave length of a photon of energy equal to the rest energy of the particle described. The paths depend on the relative velocity between the described "entity" and the observer, and represent the De Broglie wave. The merits of the proposed model are summarized and its role in relation to the established description by quantum mechanics discussed. It is concluded that it supports the existence of the proposed (EZBW), and offers a description of quantum behaviour without quantum mechanics.

\* Retired Professor of physics, Utrecht University, The Netherlands

Goethestr. 60, 79100 Freiburg, Germany

e-mail: aniehaus@outlook.de

1 .Introduction

The behaviour of electrons in the nonrelativistic regime is correctly described by the Pauli-Schrödinger theory, into which the spin is "ad hoc" introduced in the form of spinors with non-classical properties [see, f. i. [1]].

In a recent publication [2], we have shown that, the behaviour of electrons subjected to spin-measurements can also correctly be described in terms of a classical statistical model which uses a probability-density distribution (PDD) of the directions of an "instantaneous angular momentum" that changes during an assumed fast, periodic, "Zitterbewgung" (ZBW). A (ZBW) that would allow a changing instantaneous angular momentum is not predicted by the Dirac equation, while (ZBW) as a qualitative concept of interpretation of the Dirac equation exists since the beginning of quantum mechanics [3, 4, 5 ]. An "extended" (EZBW), whose existence is the basis of our statistical description in terms of a (PDD), has only recently been proposed as a hypothesis in theoretical analyses of the Dirac equation [6], and in model descriptions of the electron [7, 8, 9].The validity of these models, however, has never been established. No experimental evidence of an (EZBW) has been reported so far.

Our statistical model, therefore, which is *based* on the existence of an (EZBW), and *explains* established experimental data, constitutes important support for an (EZBW) and a corresponding substructure of the electron, if it is valid.

In the present paper we report results of our attempt to further develop the model, and to support its validity. We proceed in the following way:

(i) In the next paragraph we construct the angular dependence of the length of the "instantaneous position vector", and the probability density of its direction, from the corresponding value of the instantaneous angular momentum and its (PDD), obtained in [2].

(ii) In paragraph 3, we construct closed curves of "instantaneous positions" in real space, which are consistent with the distributions obtained in [2], and which reproduce all experimentally obtainable quantities as curve-averages. These closed curves represent individual particles.

(iii) In paragraph 4, we introduce proper time to parameterize the angles ($\vartheta$, and $\phi$), and in this way obtain paths representing the motion for different velocities relative to the observer: the (EZBW). All relevant experimentally accessible properties of the electron are in this way explained as resulting from the (EZBW).

(iv) In paragraph 5, we summarize the results, and discuss some of the many questions that arise.

## 2. The position vector

The probability density of the directions of the instantaneous angular momentum, in the (+) - state in a context defined by a magnetic field in direction (Z), was found in [2] to be given by the function

$$\text{PDD }(\vartheta,\phi)=(dN/d\omega)=\cos(\vartheta)/\pi, \tag{1}$$

with ($\vartheta$) the polar angle, and ($\phi$) the azimuth angle. The integrand $dN=(PDD)(\omega)\,d\omega$ is the probability that the instantaneous angular momentum vector (**il**($\vartheta,\phi$)) during the period of an (EZBW) points into a differential surface area $d\omega=\sin(\vartheta)d\vartheta d\phi$ of the unit sphere. A cut through the (PDD) in a plane containing the (Z)-axis is shown in **Figure 1**.

**Figure 1**: *The (PDD) of the instantaneous angular momentum for a (+)-state in the context defined by the (Z)-axis*

As indicated in the figure, the probability density vector can be decomposed into two vectors of *constant* length ($1/(2\pi)$), one pointing into the Z-direction, and the other being a radius-vector of the sphere around the point (0, 0, z=$1/(2\pi)$).

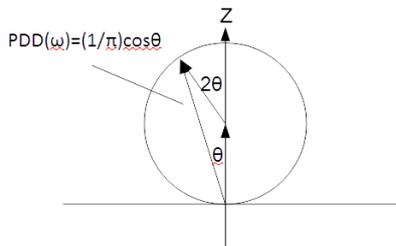

For a given functional form of the length of the instantaneous angular momentum vector **il** ($\vartheta$, $\phi$), averages over the period of the (EZBW) can be calculated from the (PDD). As shown in our previous paper [2], assumption of the functional form

$$|\mathbf{il}(\omega)|=\hbar\cos(\vartheta), \tag{2}$$

yields the average angular momentum vector **S**=(0, 0, ℏ/2), and the average projection onto the (Z)-axis $s_z$=ℏ/2, i.e. the experimentally determined spin properties.

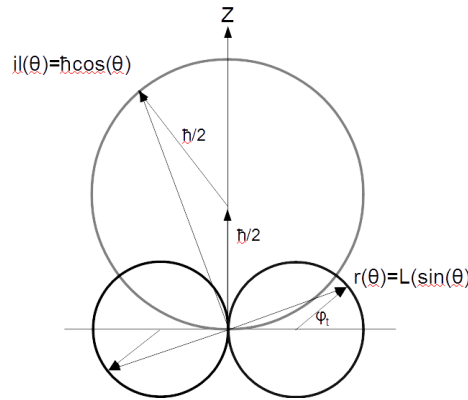

Using this information on the instantaneous angular momentum **il** (ϑ, ϕ), we now construct a vector of possible instantaneous positions, **r** (ϑ, ϕ). The two vectors are connected via relation

**il** (ϑ,ϕ)=**r** (ϑ,ϕ) × **p** (ϑ,ϕ) ;  |**il** (ϑ,ϕ)| = ℏ cos(ϑ),  (3)

where **p**(ϑ,ϕ) is an instantaneous momentum vector during the (EZBW). As an extension of our model we now *assume* that the vector **p**(ϑ,ϕ) is independent of (ϑ), and has the direction of the normal of the instantaneous plane defined by **r**(ϑ,ϕ) and the (Z)-axis. Relation (3) then yields:

|**r** (ϑ, ϕ)| =L sin(ϑ), L=ℏ/|**p**|

(4)

The conditions characterising our extended model are represented in **Figure 2**:

The angular probability density of the instantaneous position vector **r** (ϑ,ϕ), which we will call (PDDP), is determined by the (PDD) of the instantaneous angular momentum vector by the requirement that 2(PDDP) sin($ϑ_p$) d($ϑ_p$) = (PDD) sin($ϑ_l$) d($ϑ_l$). With relation (1), and taking into account that sin($ϑ_p$)=|cos($ϑ_l$)| (see Fig 2), this leads to the distribution

(PDDP)(ϑ,ϕ) = |cos(ϑ)|/(2π) . (ϑ = 0…..π)  (5)

**Figure 2**: *Polar plot showing the relation between instantaneous position r(ϑ,φ), and instantaneous angular momentum il(ϑ,φ) according to our model.*

The surface of the torus, on which the possible instantaneous positions are located, is shown in **figure 3 a**, and the probability density of directions of the position vector (PDDP), given by relation (5), is shown in **Figure 3 b**.

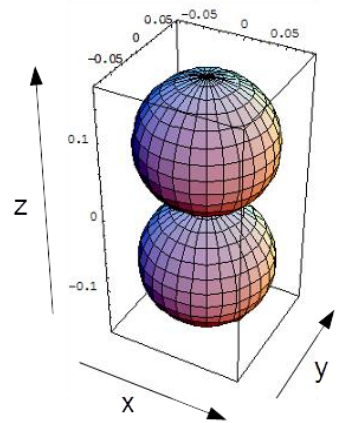

(a)

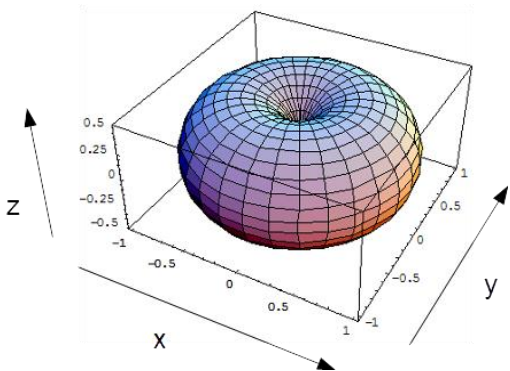

**Figure 3 :** (**a**) *The torus of possible locations in units of L=ℏ/p, and* (**b**) *the directional probability density distribution of the position vector(PDDP).*

The distributions shown in figures 3 a, and 3 b, together with relation (3) and the assumption made for the momentum (**p**), constitute our model. Averages <f(**r**)> over the period of a fast (EZBW), i.e. measurable properties of the electron, can be calculated as average over the instantaneous positions:

<f(**r**)> = ∫∫ PDDP(ϑ,φ) f(**r**) sin(ϑ) dϑ dφ = (1/(2π))∫∫ f(**r**) │cos(ϑ)│ sin(ϑ) dϑ dφ   (ϑ=0…π), (φ=0…2π)

(6)

For all properties considered so far, we obtain in this way the established experimental results, i.e. the results also predicted by quantum mechanics.

$$< \mathbf{il} > = \hbar/2 \, \{0, 0, 1\} = \mathbf{S} \text{ (Spin)} ; \tag{7 a}$$

$$< |\mathbf{il}| \sin(\vartheta) > = \hbar/2 = s_z \text{ (spin projection)}; \tag{7 b}$$

$$< \mathbf{r} > = 0; \tag{7 c}$$

$$< |\mathbf{r}| \sin(\vartheta) > = \hbar/(2\,p) = L/2 = r_c; \tag{7 d}$$

$$< \pi\, r^2 > = (\pi/2)(\hbar/p)^2 = \pi\, L^2/2 = <A> \text{ ( av. area)} \tag{7 e}$$

If the elementary charge (e) is ascribed to the position (r), a circular current (I) around the Z-axis of magnitude I=e c/(2π$r_c$) arises, and defines an instantaneous magnetic moment im=I A = π$r^2$e c/(2π$r_c$) . With result (7e) and (7 d), we thus obtain the average magnetic moment:

$$\mu = <im> = e\, c/(2\pi r_c) <\pi r^2> = e\, c\, \hbar/(2p) = \mu_B , \tag{8}$$

with ($\mu_B$) being the Bohr magneton if the momentum is replaced by (m c ), with (m) the rest mass of the electron. We see that, the model predicts the correct magnetic moment – including the "anomalous" g=2 factor – in a classical way.

## 3. Paths of instantaneous positions

Paths of instantaneous positions are obtained if the angles ($\vartheta$, and $\phi$) in relation (6) depend on each other. The integrand in (6) suggests, that the relevant averages <f(**r**)> obtained by integration over the solid angle sin($\vartheta$)d$\vartheta$d$\phi$ (see 7, 8), can also be obtained as average over such paths. The condition is that, the dependence between ($\vartheta$) and ($\phi$) is linear, and the paths are closed. Closed paths arise if the ratio of the angles is a natural number (n). To describe paths, we introduce the torus angle $\phi_t$=2$\vartheta$ (see **Fig.2**), and the angle on the circle $\phi_c$=$\phi$. The linear dependence we describe as $\phi_c$=$n_t$ $\phi_t$ if $\phi_t$< $\phi_c$, and as $\phi_t$=$n_c\phi_c$, if $\phi_c$<$\phi_t$. In this way closed paths are characterized by the natural numbers $n_c$ and $n_t$ that can vary as ( $n_c$, $n_t$ = ±1, ±2, ±3, ……..±∞), and by a variation of the angles as ( $\phi_c$= 0….2π, $\phi_t$=0….4π) for the chosen path. An explicit example of the position vector, using relation (4), for a path characterized by ($\phi_t$< $\phi_c$) is given in (9) below. Also given in (9) is the momentum vector **p**($\phi_t$, $n_t$)

**r**($\phi_t$, $n_t$) = (x,y,z),  **p**($\phi_t$, $n_t$)=($p_x$, $p_y$, $p_z$)

$x = L \cos^2(\phi_t/2) \cos(n_t \phi_t)$,

$y = L \cos^2(\phi_t/2) \sin(n_t \phi_t)$,     $\phi_t = 0 \ldots 4\pi$                                      (9)

$z = (L/2) \sin(\phi_t)$,

$p_x = -\hbar/L \sin(n_t \phi_t)$,   $p_y = \hbar/L \cos(n_t \phi_t)$,   $p_z = 0$

With relations (9), paths of the position vector, and of the angular momentum vector, can be calculated for various conditions. **Figures 4a,b** show, as examples, a 3D-plot of the path of instantaneous positions for $n_t = 10$, and a 3D-plot of the path of the instantaneous positions for $n_c = 10$.

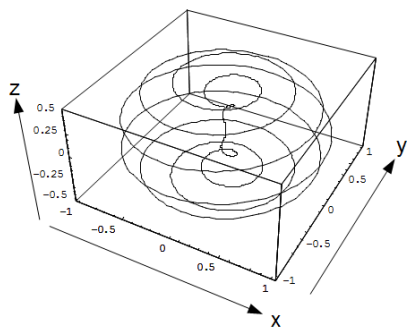 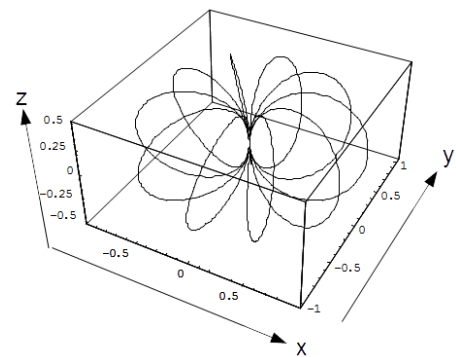

                               (**a**)                                                                      (**b**)

**Figur 4 a, b**: *Paths of the instantaneous positions in units of $L = \hbar/p$. (**a**) for $n_t = 10$, and (**b**) for $n_c = 10$.*

3D plots of the paths of the *instantaneous angular momentum vector*, are shown in **Figures 4c,d**. (**c**) for $n_t = 10$, and (**d**) for $n_c = 10$.

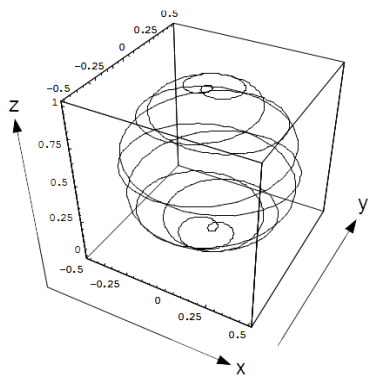 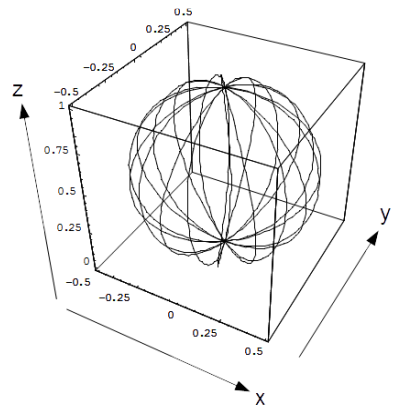

(**c**)                                                                                    (**d**)

**Figure 4c,d**: *The path of the instantaneous angular momentum vector in units of ℏ for the cases $n_t=10$ (c), and $n_c=10$ (d)*

All averages obtained using the spatial distribution of the instantaneous position vector (see results (7a-7e), and result (8)), can now be calculated as averages over the respective paths. It turns out that, results identical to the results (7a-7e, and (8)) are obtained for all allowed paths, i.e. for ($n_c$, $n_t = \pm 1, \pm 2, \pm 3, \ldots \pm \infty$). We conclude from this that, *individual particles* characterized by the paths own the corresponding properties at different conditions, and also, that the *distribution* of instantaneous positions constructed in paragraph 2, corresponds to an *average* over paths of *individual particles* with different initial conditions. These results suggest strongly that, the distinction of paths by different ($n_t$, $n_c$)-values corresponds to the necessary difference of the description of particles at different relative velocities between particle and observer, where the different velocities do not influence the observed properties. The relation between different ($n_t$, $n_c$)-values and the corresponding different relative velocities is described in the next paragraph.

## 4. Time dependent paths of the position vector

We introduce proper time by introducing frequencies for the circular- and for the toroidal variation of the corresponding angles, by writing $\phi_c=\omega_c t$, and $\phi_t=\omega_t t$. We consider the case ($\phi_t < \phi_c$) and $\phi_c = n_t \phi_t$, outlined in the preceding paragraph, and demonstrated in **Figure 4a** for $n_t = 10$. The rotation axes defining $\phi_c$ and $\phi_t$ are perpendicular to each other, and the radii ($r_c$) and ($r_t$) both have the value of half the reduced Compton wavelength. We define the quadratic sum of the frequencies, $\omega_s = (\omega_c^2 + \omega_t^2)^{1/2} = 2\omega_0 = \omega_s$, which characterizes the "entity". Using the relation between the frequencies defined above, this leads to $\omega_t (1+n_t^2)^{1/2} = \omega_s$, and thus to the following general relations:

$$\omega_t = 2\omega_0 (1+n_t^2)^{-1/2}, \qquad \omega_c = 2\omega_0 n_t (1+n_t^2)^{-1/2} \tag{10}$$

Introducing these frequencies into relations (9), one obtains time dependent positions $\mathbf{r}(t, n_t)$ – i.e., paths that are different for different relative velocities between the observer and the system at $<\mathbf{r}(t)>=0$ (see 7c). If the relative velocity $\mathbf{v}=\beta c \mathbf{e}_{rel}$ is chosen to have a certain direction (indicated by the unit vector $\mathbf{e}_{rel}$), this is taken into account by the corresponding change of the position coordinates, by writing $\mathbf{r}(t,\beta) = \mathbf{r}(t)+\mathbf{v}t$. If we choose a relative velocity in (Z)-direction, this leads to the following description of the system, based on relation (9):

$$\mathbf{r}(t,\beta) = (x, y, z) \qquad (11)$$

$x = L \cos^2((\omega_t t+\pi)/2) \cos(n_t \omega_t t)$, $y = L \cos^2((\omega_t t+\pi)/2) \sin(n_t \omega_t t)$, $z = (L/2)\sin(\omega_t t+\pi)+\beta c t$

The path for the internal motion is described by the position vector $\mathbf{r1}(t, \beta) = (x, y, z-\beta c t)$.

We need a relation between $(\beta)$ and $(n_t)$ in order to get velocity dependent paths from (11). Looking at relations (10), we notice that $(\omega_t)$ becomes twice the De Broglie frequency, which is given by $\omega_{DB}= \beta\omega_0 = \beta c/L$, if we chose

$$\beta = (1+n_t^2)^{-1/2} \qquad (12)$$

As an extension of the model we assume (12) to be correct, which then yields for the frequencies the relations

$$\omega_t = 2\beta c/L \text{ and } \omega_c = 2n_t\beta c/L \qquad (13)$$

With relations (11, 12, 13) we now have a rather complete description of the spin-(1/2) particle in terms of paths in real space of the "quantum" during its (EZBW). The description explains the wave-particle dualism reflected in the De Broglie frequency, which is represented by paths calculated for $\mathbf{r}(t, \beta)$ (see **Figure 5** below). It is further remarkable that, relative velocities characterized by a natural number $(n_t, n_c)$ are special, because for these velocities the time period is given by $2\pi/\omega_{t,c}$, while for numbers in between the natural numbers the period is longer. This predicts a kind of quantization of relative motion. Finally, from relation (12) it is also evident that *both types of paths* exemplified in **Figure 4** arise when $(\beta)$ varies in the physically possible region from zero to 1. In the region $(0<\beta<2^{-1/2})$, one has paths $\mathbf{r1}(t, \beta)$ of the internal motion of the type shown in **Figure 4a**, and for the region $(2^{-1/2}<\beta<1)$ the paths are of the type shown in **Figure 4b**. For all these paths, the same averages as given in (7a-7e, 8), are now obtained as time averages over a period of the corresponding (EZBW). In the limiting case (ß→0, $n_t$→∞) the frequencies become ($\omega_t$→0, $\omega_c$→2$\omega_0$), and in the case (ß→1, $n_c$→∞), the frequencies become ($\omega_t$→2$\omega_0$, $\omega_c$→0).

A parametric 3D plot of the position **r**(t, ß) in real space during one period of the (EZBW), calculated using relations (10, 11, 12) for the case $n_t$=10, is shown in **Figure 5**.

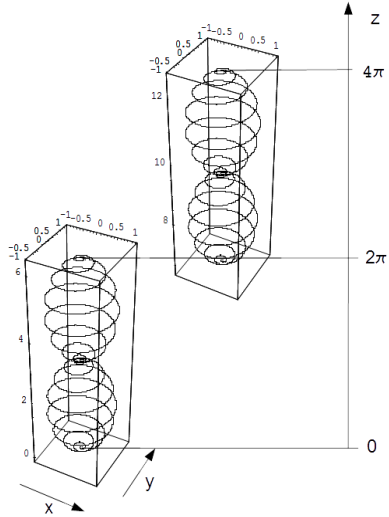

**Figure 5**: *A parametric 3D plot of the positions in real space (in units of L) during one period of the (EZBW), for the case $n_t$=10 (left figure). Path of the "entity" during the second period (right figure). Progress of the "entity" in z-direction is seen to proceed at the velocity v=2πL/τ=2πL$\omega_t$/4π= (L/2)$\omega_t$=ßc (see relation (13).*

The modulation of the lateral size of the system is due to the torus frequency $\omega_t$, which is twice the De Broglie frequency. The extension of the "entity" in z-direction during one period is 2πL, independent of (v=ßc), but its progress in z-direction as a function of time occurs at velocity (v). A thorough discussion of these paths is beyond the scope of the present paper. We expect, however, that uncertainty relations as well as interference phenomena will be describable.

## 5. Summary and discussion

We have presented a model which describes the electron in terms of paths in real space of possible positions of a "quantum" which carries out an extended periodic Zitterbewegung (EZBW). The model is completely general. The only quantity that specifies the described spin-1/2 particle, is the momentum of a photon whose energy equals the rest energy of the particle. Qualitatively, the scenario the model suggests may be summarized as follows: The "quantum", which forms the photon when it follows a straight axis and has momentum (p=mc) in direction of that axis, represents the particle of mass (m) when its axis forms a circle around a fixed point in space and is thus completely localized. Its possible positions then lie on a torus around the fixed point, with the torus radius being equal to the radius of the circle the axis forms around the fixed point. Examples of paths are shown in **Figures 4**. The energy of this electromagnetic "entity", which has a size equal to the reduced Compton wave length $\hbar/mc$, is $E=mc^2$, with (m) being the relativistic mass. In paragraph 4, proper time is introduced, which leads to the description of paths in terms of frequencies for toroidal- and circular variation of instantaneous positions. The variation of these frequencies with relative velocity between observer and "entity" completes the model. The toroidal frequency turns out to be equal to twice the De Broglie frequency, and the quadratic sum of the two frequencies is constant and equals twice the frequency of the "free" photon that has the same energy as the "entity".

Thus, the model implies a mechanism that describes the electrodynamic origin of mass, and in this way "explains" the equivalence relation $\hbar\omega=mc^2$.

The following observable properties of the "entity" - which is to be identified with the free electron - are obtained as averages over an (EZBW) by elementary calculations, and are found to agree with experiment.

(i) Spin of $\hbar/2$ is obtained as average of angular momentum of the quantum during a period of the (EZBW). Also spin projection of $\hbar/2$ is obtained as average of angular momentum projection during the period.

(ii) If the elementary charge (e) is ascribed to the position of the quantum, the magnetic moment of the free electron is predicted to equal the experimental value of one Bohr magneton. No "ad hoc" introduction of a g=2 factor is necessary.

(iii) The De Broglie frequency is identified as half the torus frequency. In this way, the "wave particle duality" of the electron is explained. The factor of two accounts for the fact that the De Broglie frequency describes the probability *amplitude*, while the torus frequency describes the *probability*.

(iv) The relativistic mass- and energy variation with relative velocity is automatically taken into account by the corresponding variation of $L=(\hbar/p)$, and of the frequencies describing the "entity".

The results above support the validity of the model, which therefore offers an alternative description to quantum mechanics, at least for the phenomena considered.

There arise, of course, many questions concerning the role of the presented model. Below we discuss the most obvious ones.

First, what is the relation between the model and non-relativistic quantum mechanics? Since the model predicts the correct magnetic moment as an average over an (EZBW)-period, at any relative velocity, and in addition predicts the same frequency for the (EZBW) as the Dirac equation does for the (ZBW), we conclude that it describes the Dirac particle, also in the non-relativistic region, in contrast to the Pauli-Schrödinger theory. Further, the phases of wave functions correspond to phases of change of the possible positions of the "quantum" in the model. For instance, the relative phase appearing in the singlet state wave function between the wave functions of the two electrons, is reflected in the phase-locked paths of the type shown in **Figure 5**, for two electrons of opposite spin, moving in opposite directions, and having a common origin. An (EPR)-paradox does not arise.

Secondly, what does the "quantization" of relative velocity, implied in relation (12), mean? As shown (see 7c), the instantaneous electric dipole moment <e**r**> - present during the (EZBW) - averages to zero over a full period. For velocities $v=\beta c=c\,(1+n^2)^{-1/2}$ which do not belong to a natural number (n), the period can be substantially longer than the one determined by (n), and an average electric moment persists until the longer period is completed. Also the average angular momentum vector – the spin – has x- and y-components until the full period is completed. Since the average electric dipole moment may lead to interactions, the translational motion at relative velocities belonging to natural numbers (n) may be regarded as especially stable. Since, during an acceleration of electrons the velocity varies continuously through regions *not* belonging to natural numbers (n), one may speculate that the observed radiation during acceleration may be explained by such incompletely averaged electric moments.

Further, the question of antimatter – the positron –we did not mention. Qualitatively, we argue as follows. The model uses two frequencies (see relations 10, 11), which can have positive or negative sign. There are four combinations of signs: (+, +), (-, -), (+, -), (-, +). The first two correspond to positive- and the second two to negative polarization of the circulating photon, and therefore are different

"entities" and represent positron and electron. The two combinations of signs, possible for each of the particles, define their two spin-states. We did not consider the question of charge. However we would expect that the different polarizations yield opposite static charges (±e) at the center of the "entities".

Finally, since the model predicts the g=2 factor correctly, the question arises why it fails to predict the (g-2) deviation of 0.002322…Bohr magnetons? If the deviation is ascribed to self-interaction, the interaction of the magnetic moment with the calculated average electric moment due to the average distance of the charge from the rotation plane $<e|z1|> = e(1/\pi)L$ (see relation (11)), would be a possible candidate. This speculation would lead to a correction of the g- factor. Assuming the coupling constant between electric moment $<e|z1|>$, and the magnetic moment $\mu_B$, to be the fine structure constant ($\alpha$), a correction of $(1/\pi)\alpha = 0.002322…$would arise. This is the first term of the quantum-electrodynamic correction of the g=2 factor in terms of powers of ($\alpha$).

In conclusion, the demonstrated merits of the model presented strongly suggest its validity. The model supports the existence of the proposed (EZBW), and suggests the purely electromagnetic origin of mass. And last but not least, it demonstrates that microscopic phenomena can be described without quantum mechanics, and thereby "explains" paradoxes known to be connected with "understanding" quantum mechanics.